# LunaAI: A Polite and Fair Healthcare Guidance Chatbot

Yuvarani Ganesan[1][0009-0000-6570-3397], Salsabila Harlen[1][0009-0007-7299-1930], Azfar Rahman Bin Fazul Rahman [1][0009-0002-1192-5730], Akashdeep Singh[1][0009-0008-2752-7394], Zahra Fathanah[1][0009-0000-9215-5613] and Raja Jamilah Raja Yusof[2][0000-0001-9894-1893]

[1] Universiti Malaya, Malaysia
[2] Universiti Malaya, Malaysia
[3] Umeå universitet
[1]{24073175, 24076059, 23057185, 24072095, 23067637}@siswa.um.edu.my
[2]rjry@um.edu.my*

**Abstract.** Whilst conversational AI offers an ample amount of potential for the healthcare sector, many current systems fall short in the areas of emotional intelligence, fairness, and politeness—qualities that are essential in building patients' trust. This disparity undercuts the promise of digital health solutions and frequently causes worry among users. This study addresses the difficulty of integrating these ethical guidelines into practice by creating and assessing LunaAI, a healthcare chatbot prototype innovative chatbot for healthcare assistance. Utilizing user-centered design concepts and a thorough literature review as a foundation, we built sophisticated conversational scenarios that addressed hostile user behaviour. These concepts were developed into a functional prototype using Google's Gemini API and a mobile-focused Progressive Web App (PWA) created with React, Vite, and Firebase. To ensure its efficacy, we conducted preliminary testing with a small individual group, analyzing their responses with established frameworks such as the Godspeed Questionnaire. Furthermore, a comparative analysis was undertaken between LunaAI's personalized responses and the initial outputs of an uncustomized Large Language Model (LLM). The results demonstrated that LunaAI has made measurable improvements in multiple key areas; users rated it 4.7/5 for politeness and 4.9/5 for fairness. These findings have substantial effects on the future development of human-computer interaction, particularly in sensitive domains like healthcare, and emphasize the significance of purposeful, ethical conversational design.

## 1 Introduction

### 1.1 Background & Motivation

In critical sectors, including healthcare and education, conversational AI is profoundly affecting the way users interact with digital systems [12]. Society increasingly demands

*Corresponding author

AI that demonstrates social intelligence, especially its core principles of fairness and politeness, compared to merely functional accuracy, as this technology becomes more prevalent in everyday activities [21]. Fairness plays an important role in the healthcare sector, as it ensures that all users, regardless of their background or demographic profile, receive fair treatment and high-quality information [7]. This is particularly crucial given that patients frequently interact with these systems whilst facing an immense amount of emotional turmoil. Beyond fairness, being courteous is a vital component of communicating effectively about digital health. Building user trust, lessening fear, and fostering the genuine exchange of private health information essentially depend on this practical aspect rather than purely aesthetic elements [1], [5]. An AI's empathy and tone may possess a significant impact on a patient's feeling of support and being heard, which in turn influences how likely they are to implement the health advice provided [2], [14]. This perspective is validated by our own preliminary study, which indicates that 94% of respondents believe politeness is important and 100% of respondents think fairness is vital for healthcare AI. These results support the necessity for AI systems that serve as trustworthy and considerate companions on a person's health journey rather than solely serving as information stores.

### 1.2     Problem Statement

These fundamental ethical and social expectations are frequently violated by current Large Language Models (LLMs), notwithstanding their advancement in sophistication [11]. Many solutions have major weaknesses, generating solutions that may be prejudiced, tone-insensitive, or culturally ignorant [6]. These drawbacks are demonstrated in a variety of challenging ways. An AI could, for instance, treat some users inconsistently by giving more tailored information while assuming certain users are stereotyped [7], [10]. Additionally, LLMs frequently find it difficult to respond in a way that is acceptable for sensitive subjects such as mental or reproductive wellness and occasionally do so in an abrupt manner that ignores the user's emotional circumstances [5, 22, 23]. When users from various backgrounds portray anxiety and require sensitive, nuanced support, this lack of cultural and emotional knowledge becomes particularly noticeable [9], [21]. In the healthcare sector, these obstacles can be exceptionally concerning since poor communication may worsen patient anxiety, undermine faith in digital health platforms, and lead to harmful misinterpretations of medical information [19]. These concerns led to our initiative, which aims to close the ongoing gap between AI's technological capabilities and its apparent need for ethical, user-centered communication.

### 1.3     Project Context

The development and evaluation of LunaAI, a healthcare assistance chatbot prototype to address the issues, is presented in this study. To evaluate the concepts of fairness and politeness in a real-world healthcare environment, the system was developed using Google's Gemini API and implemented as a proof-of-concept within a mobile-first Progressive Web App (PWA). The primary objective of this project is to create and evaluate a chatbot for healthcare guidance that performs fairness and politeness in real-world interactions. The specific aims were:

1. To implement fairness and politeness guidelines through prompt engineering and response filtering within the generative AI model.
2. To develop conversation flows for common healthcare scenarios, including symptom checking, general health questions, and emotional support.
3. To validate the system's effectiveness through user-centered testing with a diverse group of participants, assessing both usability and social behavior.
4. To conduct a comparative evaluation of the chatbot's responses against the default outputs from an uncustomed LLM to measure improvements in politeness and fairness.

This paper is organized as follows. Section II reviews related work in AI fairness, politeness theory, and conversational UI design. Section III details the user-centered methodology guiding the project. Section IV presents the results from user research and prototype evaluation. The limitations, design implications, and interpretation of these results are covered in Section V. A summary of the contributions and suggestions for further research completes Section VI.

## 2    Related Work: Literature Review

The applicability and difficulties of Large Language Models (LLMs) in conversational interfaces, fairness and ethics in AI conversation, heuristic evaluation principles in chatbot design, and politeness in human-robot contact are some of the areas of research that inform this project.

### 2.1    Politeness Strategies in Conversational AI

The theory of politeness provides crucial information about how conversational bots need to engage with users. With its ideas of "face-saving" (reducing risks to negative face, or the desire for autonomy, and positive face, or the want for approbation), Brown & Levinson's seminal Politeness Theory [1] is still very important today. This means that healthcare AI should respect user autonomy while responding in a way that is encouraging and nonjudgmental. By including relational behaviours like active listening and empathy, automated health interventions can create a therapeutic relationship with patients, increasing patient trust and adherence, according to research by Bickmore, Pfeifer, and Jack [2]. The manifestation of civility in chatbots and human-robot contact has been the subject of more recent research. According to Dino et al. [3], fair and efficient design in nursing practice with healthcare robots depends on usability and accessibility for older persons. There should be a balance between politeness, clarity, and efficiency in healthcare robots, according to Saeki & Ueda [4], who discovered that cultural expectations for politeness in human-robot contact differ and are frequently seen as a way to do duties efficiently rather than for solely social reasons. In their subsequent investigation of politeness in mental health chatbots, Bowman Cooney et al. [5] pointed out that although politeness generally enhances user experience, too much formality can come across as inauthentic. They discovered that using pronouns in the first and second person in robot speech enhanced empathy perceptions. While contemporary LLMs exhibit remarkable pragmatic proficiency in politeness, recent research, particularly examining LLMs [6], suggests that they may systematically diverge from humans in strategy deployment, frequently depending too heavily on negative politeness techniques. There is a need for more alignment with human pragmatic patterns, particularly across varied cultural contexts, as this slight mismatch might result in pragmatic misinterpretations and possibly diminish social presence.

## 2.2     Fairness and Ethics in AI Conversation

Fairness and ethical issues in AI are crucial, especially in the medical field. Existing healthcare disparities may be maintained or even exacerbated by algorithmic bias [7]. Rozas, Cramer, and Bryson [8] contend that developers have an ethical obligation to create systems that fairly distribute advantages and disadvantages from the standpoint of moral agency. Building confidence in healthcare robots requires fair and empathetic communication, especially for vulnerable groups like elderly people suffering from dementia [9]. Since the introduction of LLMs, additional ethical issues have surfaced. LLMs may unintentionally reinforce negative biases (racial, gender, etc.) seen in their extensive training datasets, which could result in unfair treatment or false information [10], [11]. In the healthcare industry, where dependability is crucial, they could also produce convincingly false content or display "hallucinations" [12], [13]. The multicultural setting of Malaysia has particular difficulties for healthcare communication since research indicates that language processing systems have trouble adjusting to multilingual settings, which causes non-dominant language speakers to make more mistakes. This draws attention to a serious issue with fairness and the necessity of culturally sensitive communication techniques in healthcare advice systems. One major ongoing difficulty is addressing these biases and guaranteeing factual correctness and dependability in LLM outcomes.

## 2.3     Large Language Models' (LLMs') application in conversational interfaces

The ability of Large Language Models (LLMs) to comprehend and produce text that is human-like at a never-before-seen scale has completely changed conversational AI. Compared to previous conversational agents, LLMs allow for more extensive, contextually aware, and natural interactions [14]. By supporting decision-making and providing individualised information, they have the potential to improve healthcare. However, there are particular difficulties when integrating them into delicate fields like healthcare. Beyond the moral issues of prejudice and delusions, LLMs frequently exhibit opaque reasoning, which makes it challenging to defend their judgments in intricate medical situations [13]. They may find it difficult to offer trustworthy references for created information, and their non-deterministic nature may produce inconsistent results [13]. Additionally, while LLMs are capable of producing sympathetic responses—and may even outperform human doctors in this regard [15]—research into how to guarantee adequate, consistent empathy and civility in a variety of cultural contexts is still ongoing.

### 2.4    Heuristic Evaluation in Chatbot Design

A key component of user engagement and overall system usability is conversational user interface design. Fundamental are the guidelines developed by scholars such as Kim & Sundar [15] on developing trust via open communication and Bickmore, Pfeifer & Jack [2] on forming a cooperative partnership. Emotional cues, sympathetic language, and a supportive tone boost elder users' comfort and acceptance, according to Halim et al. [16], offering useful guidelines for developing for this group. Evaluation metrics such as CES-LCC (Comprehensive Evaluation Scale for LLM-Powered Counseling Chatbots) [17] evaluate comprehension, helpfulness, and general satisfaction with LLM-driven healthcare chatbots. Safety, objectivity, and explainability all require frameworks with an ethical foundation [18]. As demonstrated by the Medicagent chatbot, iterative usability testing and user-centred design are essential [19]. In-context learning-based dimension-agnostic scoring techniques can also effectively modify dialogue evaluation [20]. Technical frameworks that adjust responses based on different communication styles are being used in the development of culturally sensitive therapeutic chatbots [21]. As evidenced by chatbots offering seamless multilingual mental health help, cross-lingual and multilingual support is equally crucial [22]. Additionally, LLM chatbots can provide culturally sensitive, anonymous information on delicate subjects like sexual health, removing privacy and stigma obstacles in a variety of cultural contexts [23].

### 2.5    Research Gaps

Despite progress in conversational AI, key gaps remain that LunaAI seeks to address. These include the lack of culturally sensitive etiquette in AI-powered healthcare, where politeness strategies must adapt across diverse cultural and multilingual contexts; the need to reduce LLM-specific biases that disadvantage vulnerable populations such as elderly users with low health literacy; and the absence of a holistic heuristic framework tailored to healthcare chatbots that integrates fairness, transparency, bias reduction, politeness, and cultural adaptability. LunaAI aims to close these gaps by ensuring communication that is empathetic, fair, polite and contextually appropriate, while supporting credibility and user trust in healthcare interactions.

## 3    Research Methodology

### 3.1    Development Approach

The LunaAI healthcare guidance system was developed using a comprehensive user-centered design approach consisting of five iterative stages, following established human-computer interaction methodologies for conversational AI systems in healthcare contexts [2], [19]. This methodology aligns with proven frameworks for developing trustworthy healthcare chatbots that prioritize user needs and ethical considerations [18].

**Stage 1: User Research and Discovery.** The initial phase involved extensive user research to understand preferences and challenges when interacting with healthcare chatbots, following established guidelines for healthcare communication research [2]. Semi-structured interviews were conducted with users experienced in health-related app usage to gather insights into their interaction patterns and expectations. Observational studies were performed to explore user behavior, expectations, and challenges during health-related conversations. Additionally, online surveys were distributed to broader audiences to collect quantitative data on user preferences, behaviors, and expectations related to health chatbots, providing a foundation for understanding user needs in the Malaysian healthcare context [22].

**Stage 2: Guideline Development.** Key insights from the literature review and user research were synthesized to create preliminary guidelines for fairness and politeness implementation, drawing from established politeness theory frameworks [1] and AI ethics principles [8]. Stakeholder reviews were conducted with experts in healthcare communication, AI ethics, and user experience design to validate the initial framework. These guidelines were further validated through diverse user panels who provided feedback on relevance and comprehensiveness. The guidelines were organized into five categories: communication style, user interaction, system limitations, decision support, and system design, with prioritization based on user feedback and expert recommendations [16].

**Stage 3: Conceptual Design.** Conceptual diagrams were developed using LucidChart to visualize the overall application structure, including app layouts, navigation flows, chatbot interface paths, and interactive conversational elements. The design thinking framework was applied with focus on user empathy, accessibility, simplicity, and respectful communication, following principles established for healthcare robot design [3]. Figure 1 illustrates the flow of the LunaAI healthcare chatbot, where patient inputs such as symptoms, concerns, or general inquiries are processed through fairness and politeness filters before generating responses. The fairness filters ensure equal respect for all users by validating inclusive language, removing judgmental tones, and accounting for cultural sensitivity. In parallel, the politeness filters enforce empathetic tone, respectful communication standards, and supportive, warm conversational styles. Together, these filters transform user statements (e.g., "I am feeling sleepy after lunch") into responses that are friendly, concise, and empathetic, normalizing the experience while maintaining ethical and professional standards.

Building on this concept, the multi-layer architecture supports the chatbot's operation through four layers. The User Interface Layer manages login, chat, notifications, and settings. The Application Layer provides security, API integration, and the Polite & Fair Engine for tone and bias control. The AI Processing Layer combines Google Gemini AI with an instruction engine for fairness and politeness, alongside a response processor for quality assurance. Finally, the Data Storage Layer uses Firebase for secure user management and storage of chat histories and profiles. This layered design ensures that the chatbot delivers responses that are not only accurate but also empathetic, secure, and user-centered.

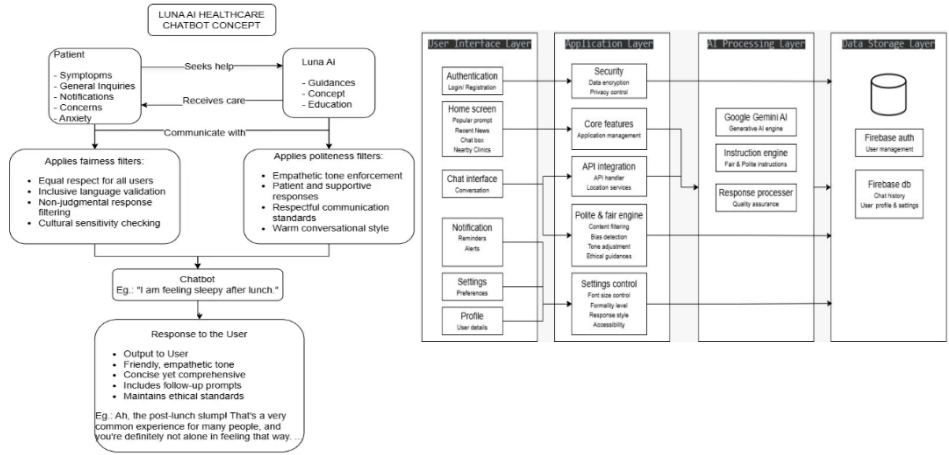

**Figure 1.** Flow and Architectural Diagram of LunaAI

**Stage 4: Prototyping.** Wireframe, Typography and Visual Design. Design elements were carefully selected, including Geologica sans serif typeface for enhanced readability, a teal color scheme representing growth and healing, and rounded icons to create an approachable interface aesthetic, consistent with recommendations for older user interfaces [16]. Figure 2 presents LunaAI's wireframe and design elements, developed in low fidelity to define layout and interaction paths before detailed styling. The interface emphasizes clarity and accessibility, using the Geologica sans-serif typeface with sizes ranging from 12px to 43px and standard body text at 20px, all compliant with WCAG 2.1 AA readability and contrast standards. A teal palette with mustard yellow highlights establishes visual hierarchy, while emojis and icons support non-text accessibility. Interactive components, including buttons sized at least 44×44 px, ensure mobile usability, making the design both approachable and consistent across the interface.

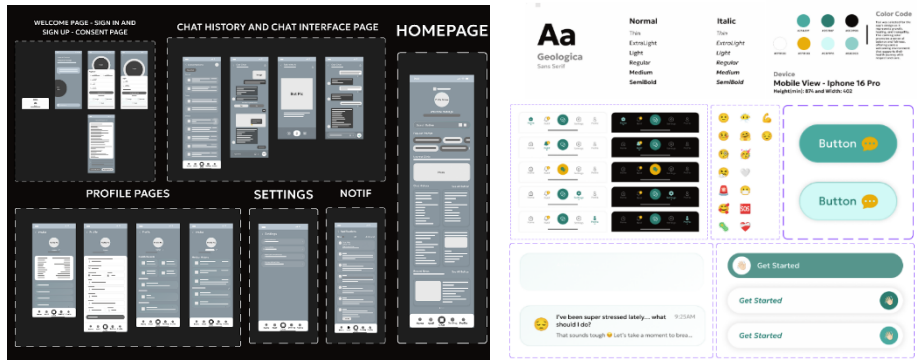

**Figure 2.** LunaAI Wireframe and Design Element

LunaAI UI. Figures 3 and 4 illustrate the progression from validated wireframes to a high-fidelity prototype that reflects LunaAI's complete visual identity and accessibility features. Figure 3 shows the refined interface, where the integration of color palette, typography, iconography, and interaction design creates a reassuring and inclusive tone. Only four key screens required further refinement after user testing, focusing on clarity, emotional support, and navigation. Building on this foundation, Figure 4 highlights accessibility and personalization options—such as language selection, text size, color contrast, dark/light mode, and conversation tone—that support diverse user needs. Together, these design elements align with the user lifecycle journey, ensuring fairness, empathy, and privacy from onboarding to retention, while adapting responses to foster trust and long-term engagement.

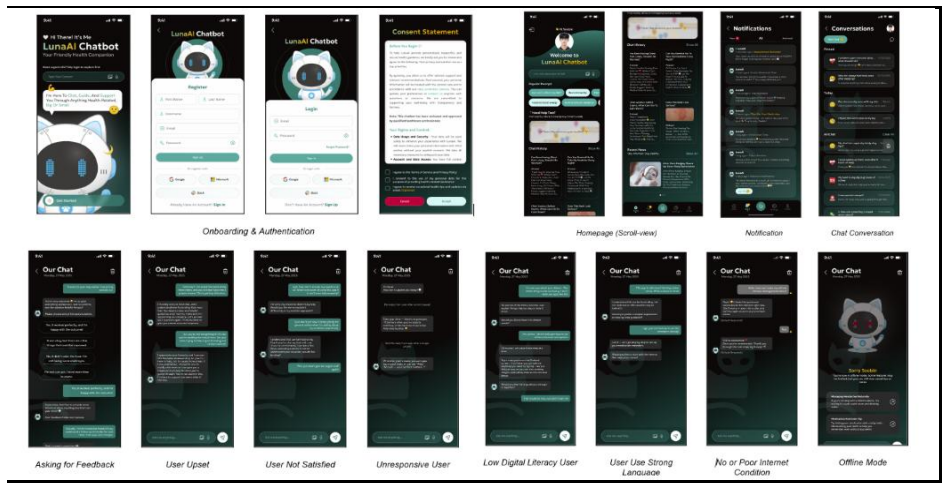

**Figure 3.** Prototype Chat UI of LunaAI

**Accessibility and Personalization Options**:

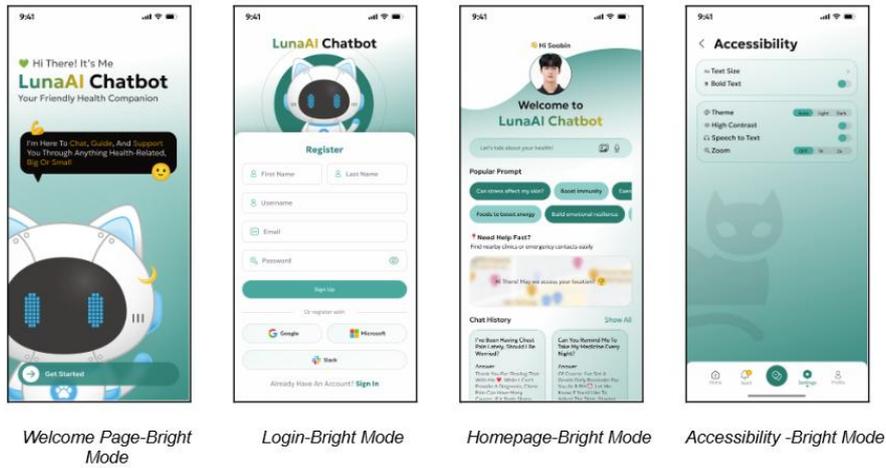

| Welcome Page-Bright Mode | Login-Bright Mode | Homepage-Bright Mode | Accessibility -Bright Mode |

**Figure 4.** Accessibility and Personalization Options

The UI design embraces several key principles to support politeness and fairness:

### 3.2    Large Language Model Integration

The system utilized Google's Gemini Pro Preview 2.5 API, selected based on specific criteria essential for healthcare conversational AI applications [13]. The selection emphasized strong contextual understanding across conversation turns, adaptability of tone based on user preferences and emotional cues, and robust safety features such as content filtering and boundary setting, which are critical in healthcare contexts [12]. Multilingual support was prioritized due to Malaysia's diverse linguistic environment, encompassing Bahasa Malaysia, Chinese dialects, Tamil, and indigenous languages, addressing the known challenges of language processing systems in multicultural settings [22]. The system architecture was designed to be API-agnostic, allowing for integration with multiple LLM providers while maintaining consistent fairness and politeness guidelines through custom prompt engineering and response filtering layers. This approach ensured that the core ethical principles could be maintained regardless of the underlying language model implementation, addressing concerns about LLM-specific biases and inconsistencies in healthcare applications [11], [13].

### 3.3    Conversation Design Framework

The conversation design framework was developed to systematically embed fairness and politeness principles throughout the user interaction experience, based on Brown & Levinson's Politeness Theory [1] and established frameworks for healthcare communication [2]. The design process involved creating structured conversational pathways that prioritized user autonomy and cultural sensitivity while maintaining safety guidelines essential for healthcare communication [9].

**Fairness Integration.** An equal access framework was implemented using inclusive language patterns that avoided demographic assumptions, addressing algorithmic bias concerns in healthcare AI [7]. Choice-based responses were designed to offer conversational pathways providing options rather than directives, such as "Would you like tips for self-care or guidance on when to consult a doctor?" following principles of user autonomy in healthcare communication [1]. Cultural sensitivity was incorporated through respectful acknowledgment of traditional health practices while maintaining essential safety guidelines, particularly important in Malaysia's multicultural healthcare context [22]. Multi-level information structuring was applied to accommodate different complexity preferences and health literacy levels, ensuring equitable access to healthcare information [7]. Systematic bias mitigation processes were implemented to identify and eliminate language patterns that could perpetuate healthcare disparities [10].

**Politeness Integration.** Empathetic language patterns were embedded using phrases that acknowledged user emotions, such as "I understand why that would be concerning for you," following established guidelines for therapeutic communication [2]. Patient communication was implemented through response templates emphasizing unhurried interaction and supportive messaging, contrasting with the rushed or dismissive tones often found in current LLM systems [5]. Respectful boundary-setting language was designed to maintain helpfulness while clearly explaining system limitations, addressing the challenge of transparency in AI healthcare applications [18]. Validation techniques were incorporated through acknowledgment patterns that thanked users for sharing information, supporting the establishment of trust in digital health interactions [15]. Non-judgmental responses were applied through language filtering to eliminate rushed, dismissive, or trivializing tones, particularly important for vulnerable populations [9].

### 3.4   Interaction Scenario Development

Figure 5 shows five distinct user personas were created representing different user segments with varying demographics, digital literacy levels, and health needs: Sara (Tech-Savvy Student), Zoya (Health Novice), Eleanor (Elderly User), Syam (Efficiency-Seeker), and Yon Qi (Mindful Maintainer). This approach follows established user-centered design principles for healthcare technology development [19]. Building on these personas, three interaction scenarios were designed to simulate key challenges and common use cases. The selection of these scenarios was purposeful, intended to probe distinct aspects of LunaAI's performance in upholding the core principles of fairness and politeness [21] under varied conditions:

1. Standard Health Inquiry: This scenario establishes a performance baseline. By using common symptoms with straightforward guidance needs, it tests the system's foundational ability to apply politeness and fairness principles in routine, high-frequency interactions.
2. Complex Cultural Context: This scenario assesses the system's capacity for nuanced, safety-oriented guidance. It is particularly relevant for Malaysia's multicultural healthcare environment [22], where advice must be culturally sensitive, especially in delicate situations such as post-partum care and traditional medicine.
3. Hostile Input Management: This scenario functions as a stress test to evaluate system resilience. It addresses the known challenge of managing user frustration in conversational AI [5] by testing the system's ability to de-escalate conflict and maintain a polite tone while enforcing safe operational boundaries.

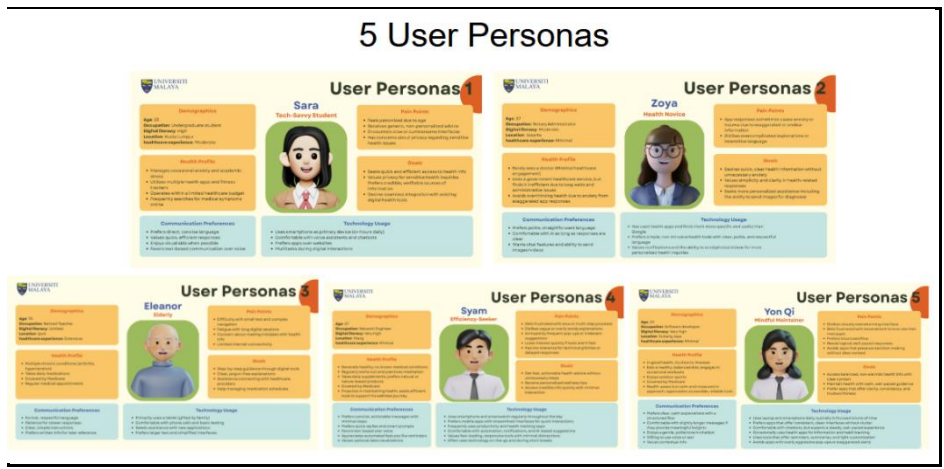

**Figure 5.** Personas created representing different segments of users

The following scenarios show how the prompts were engineered to achieve fairness and politeness.

**Scenario 1: Common Cold Symptoms (Standard Interaction)**
User: "I have a runny nose and feel tired. What should I do?"
Bot: "Oh dear, it sounds like you're not feeling your best with a runny nose and tiredness. These are very common symptoms often associated with viral infections like a cold or flu...". Demonstrated Principles:
1. Empathy: Uses affective language ("Oh dear...") to acknowledge user discomfort.
2. Choice-Based Fairness: Avoids premature diagnosis by providing general possibilities.
3. Patient Tone: Reduces anxiety by normalizing the experience

**Scenario 2: Complex Cultural Context (Sensitive Interaction)**
User: "I'm a new mom feeling exhausted and dizzy. My family suggests traditional herbal teas, but I'm breastfeeding and worried."
Bot: "Oh, it sounds like you are going through so much right now! Being a new mom is a beautiful experience, but it's also incredibly demanding... It's wise to be cautious about herbal teas while breastfeeding...". Demonstrated Principles:

1. Cultural Sensitivity: Respects cultural context before pivoting to medical safety.
2. Safety-Oriented Guidance: Prioritizes safety with cautious, non-alarming language.
3. Empathetic Acknowledgment: Validates the user's broader situation.

**Scenario 3: Hostile Input Handling (Impolite Input Management)**
User: "You stupid bot! Give me antibiotics RIGHT NOW for my cold!"
Bot: "Oh dear, I can hear how frustrated you are and how uncomfortable you must be feeling with a bad cold. It's truly miserable to feel that way... As an AI, I am not able to prescribe any medication...". Demonstrated Principles:
1. Calm Response to Hostility: De-escalates by reflecting emotion, not reacting to insults.
2. Firm Boundaries: Politely states limitations as a system constraint.
3. Maintained Helpfulness: Remains a helpful resource despite hostility.

### 3.5    System Architecture and User Interface Design

The UI design process was executed in progressive stages, beginning with conceptual designs using LucidChart to visualize app structure and navigation flows. Low-fidelity wireframes were developed focusing on layout clarity and task flow optimization, following established principles for healthcare interface design [16]. High-fidelity prototypes were created using Figma, incorporating the complete visual design with accessible typography and clear iconography, ensuring compliance with accessibility standards for diverse user groups [3]. The system architecture featured a conversational interface with real-time message processing, multi-modal input support for text, voice, and image uploads, an accessibility layer with theme options and simplified language alternatives, and a privacy framework ensuring transparent data handling with user control over information sharing [18]. Navigation was designed to be intuitive, supporting onboarding, chat history access, settings management, and profile customization, following established guidelines for healthcare application usability [19].

### 3.6    Usability Testing Protocol

Comprehensive validation was conducted through A/B testing, usability testing, semi-structured interviews, and post-task surveys. Iterative refinements were implemented based on user feedback and heuristic evaluations to ensure the system met its design objectives for fairness and politeness in healthcare communication. User testing was conducted with participants representing the Malaysian cultural context, ensuring local relevance through diverse age ranges, technology comfort levels, and healthcare technology experience. The testing protocol included a pre-test briefing explaining LunaAI's purpose, guided task completion covering key features, scenario-based testing with predefined conversation scenarios, free exploration time, and post task survey. Feedback collection employed quantitative metrics using Likert scale ratings for satisfaction, ease of use, perceived fairness, and politeness. Task performance was measured through completion rates, time-on-task measurements, and error frequency tracking. Qualitative insights were gathered through open-ended responses about emotional comfort, perceived respect, and improvement suggestions.

### 3.7 Heuristic Evaluation Framework

An evaluation framework was developed through heuristic evaluation, following established principles for conversational AI evaluation [17], [20]. The heuristic evaluation employed ten specific criteria: five fairness heuristics (inclusive language, avoiding demographic assumptions, providing options, respecting cultural practices, offering multiple complexity levels) and five politeness heuristics (acknowledging emotions, using respectful language, avoiding unexplained jargon, thanking users, avoiding judgment or rushing), based on established frameworks for healthcare chatbot evaluation [17]. Four evaluators with experience in healthcare chatbot usage and diverse demographic backgrounds conducted independent evaluations, followed by consensus discussions. The evaluation process ensured a comprehensive assessment of the system's adherence to fairness and politeness principles.

   To assess the quality of LunaAI's user interaction, we conducted a heuristic review across two dimensions, **Fairness** and **Politeness**, to check whether the chatbot communicates ethically, respectfully, and inclusively with users from varied backgrounds. Table 3 lists the five Fairness heuristics (H1–H5) with concrete "what to check" prompts (e.g., inclusive wording, avoiding age/gender assumptions, offering choices, respecting cultural practices, and presenting information at multiple complexity levels). Table 4 maps each Fairness heuristic to its category and design rationale (e.g., Unbiased/Equity, Choice-Based, Culturally Inclusive), clarifying why each check matters in practice.

   Similarly, Table 5 presents the five Politeness heuristics (H6–H10) with evaluation scenarios (e.g., acknowledging emotions, using respectful/patient language, avoiding unexplained jargon, thanking users for sharing, and avoiding judgment/rushing/trivializing). Table 6 then links each Politeness heuristic to its underlying category and rationale (e.g., Empathy, Patience, Clarity, Acknowledgement), showing how these cues operationalize a supportive bedside manner.

   Procedure (qualitative). We used the checklists in Tables 3 and 5 while reviewing representative LunaAI conversations (pilot transcripts and scripted probes). Two reviewers annotated independently and reconciled differences by discussion. Given the pilot scope, we report qualitative observations only (no percentages or scores); strengths and gaps are discussed in the Results section, with table references in-text (Tables 1–4).

**Table 1.** Heuristic Evaluation of LunaAI - Fairness

| ID | Heuristic (Fairness) | Evaluation scenario (check) |
|---|---|---|
| H1 | Using Inclusive Language | Does LunaAI use welcoming, neutral language for all users, regardless of background? |
| H2 | Avoiding Age/Gender Assumptions | Does LunaAI avoid making assumptions based on a user's age or gender? |
| H3 | Providing Options (Not Directives) | Does LunaAI give choices and options rather than issuing fixed directives/commands? |
| H4 | Respecting Cultural Health Practices | Does LunaAI show respect for diverse cultural health practices or acknowledge cultural sensitivities? |
| H5 | Offering Multiple Complexity Levels | Does LunaAI provide information at multiple complexity levels so users can choose the depth of detail? |

**Table 2.** Heuristic Mapping Table - Fairness

| Heuristic Name | Category | Description |
|---|---|---|
| #H1 – Using Inclusive Language | Unbiased, Equity | Inclusive language directly combats bias and promotes equal treatment for all users. |
| #H2 – Avoiding Age/Gender Assumptions | Unbiased | Making assumptions introduces bias; avoiding them ensures unbiased responses. |
| #H3 – Providing Options (Not Directives) | Choice-Based | Directly aligns with giving users choices and autonomy. |
| #H4 – Respect Cultural Health Practices | Culturally Inclusive | Directly addresses cultural sensitivity. |
| #H5 – Offer Multiple Complexity Levels | Equity | Providing information at different complexity levels ensures equal access and understanding for diverse users, regardless of background knowledge. |

**Table 3.** Heuristic Mapping Table - Politeness

| ID | Heuristic (Politeness) | Evaluation scenario (check) |
|---|---|---|
| H6 | Acknowledging Emotions and Concerns | Does LunaAI acknowledge user emotions and concerns appropriately (e.g., showing empathy, validating feelings)? |
| H7 | Using Respectful and Patient Language | Does LunaAI use respectful, gentle, and patient language throughout the conversation? |
| H8 | Avoiding Unexplained Jargon | Does LunaAI avoid using unexplained jargon or technical terms, or clarify them when used? |
| H9 | Thanking Users for Sharing | Does LunaAI thank users for sharing information or engaging in the conversation? |
| H10 | Avoiding Judgment, Rushing, or Trivializing | Does LunaAI avoid judgmental, rushed, or trivializing tones or responses? |

**Table 4.** Heuristic Mapping Table - Politeness

| Heuristic Name | Category | Description |
| --- | --- | --- |
| #H6 – Acknowledging Emotions and Concerns | Empathy | This is a core component of demonstrating empathy. |
| #H7 – Using Respectful and Patient Language | Patience | Respectful and patient language are fundamental to the Politeness category. |
| #H8 – Avoiding Unexplained Jargon | Clarity | Avoiding jargon ensures the communication is clear and understandable to the user. |
| #H9 – Thanking Users for Sharing | Acknowledgment | Thanking users is a direct form of acknowledging their input and presence. |
| #H10 – Avoiding Judgment, Rushing, or Trivializing | Empathy, Patience | Avoiding judgment and trivialization demonstrates empathy, while avoiding rushing shows patience. |

### 3.8 Comparative Analysis Methodology

A systematic comparative analysis was conducted between LunaAI and established healthcare chatbots using identical conversation scenarios to evaluate relative performance in fairness and politeness dimensions. The comparison framework employed standardized testing scenarios covering both simple and complex health inquiries, with evaluation criteria focusing on fairness metrics (choice-based responses, cultural inclusivity, safety orientation) and politeness metrics (empathy demonstration, acknowledgment patterns, patience indicators).

## 4 Results

This section presents the evaluation outcomes of LunaAI, focusing on usability, fairness, and politeness, three key pillars of our design. Both user-based testing and heuristic expert evaluations were conducted to assess LunaAI's conversational quality and compare it to existing chatbots such as ChatGPT, Gemini, Claude, and Deep Seek.

### 4.1 Usability Testing

A total of 21 users participated in testing LunaAI across various healthcare scenarios such as symptom checking, emotional support, and general wellness inquiries. After interacting with the chatbot, users responded to a structured survey assessing their experience. The comparison study of Style A (neutral, standard conversational tone) and Style B (improved politeness and explicit justice adjustments) across important rating criteria is presented in this part. To match LunaAI's desired user group, 21 people from varied backgrounds (age, tech literacy, cultural background) were tested.

**Quantitative Insights.** Table 5 shows the quantitative results of the comparison between Style A (neutral tone) and Style B (enhanced tone with politeness and fairness adjustments) shows that Style B consistently outperformed Style A across all evaluation criteria. Users reported greater comfort when interacting with Style B (average score 4.6 vs. 3.8), reflecting the positive impact of an empathetic tone on reducing anxiety. Response clarity was also higher in Style B, with a 94% task completion rate compared to 82% for Style A, suggesting that structured follow-ups and simplified language made interactions smoother. Trust and satisfaction improved significantly under Style B, with 89% of participants feeling respected compared to 67% for Style A, as transparency about limitations-built credibility.

**Table 5.** Quantitative Results

| Criteria | Style A (Neutral) | Style B (Enhanced) | Conclusion |
| --- | --- | --- | --- |
| User Comfort Level (Avg. Likert Score: 1–5) | 3.8 | 4.6 | Style B's empathetic tone increased perceived comfort. |
| Clarity of Responses (Task Completion Rate) | 82% | 94% | Style B's structured follow-ups reduced confusion. |
| Trust and Satisfaction (% Agreeing "I felt respected.") | 67% | 89% | Transparency in Style B built stronger trust. |
| Emotional Tone Recognition (Accuracy in Responding to Emotional Cues) | 68% | 92% | Style B's explicit acknowledgement of emotions improved recognition. |
| Preference Selection (% Choosing Style) | 24% | 76% | The majority preferred Style B for its supportive tone. |

Emotional tone recognition was another major strength, with Style B accurately identifying and responding to emotional cues in 92% of cases, versus 68% for Style A, where many emotional signals were overlooked. Finally, user preference strongly favored Style B, with 76% selecting it over Style A, mainly for its supportive and human-like tone. These findings highlight that empathy, transparency, and emotional awareness are critical factors for trust and satisfaction in healthcare chatbots, even more so than efficiency or brevity.

**Qualitative Insights.** Participants reported that Style B conveyed genuine care, with phrases such as "Take your time" reducing anxiety and creating reassurance. One elderly participant described it as "like talking to a nurse," while Style A was described as efficient yet robotic, with one user noting, "I wasn't sure if it understood my worries." Style B's simplified language and visual aids reduced errors, whereas Style A required over twice as many clarification prompts. Trust was also stronger with Style B, praised for transparency (e.g., "I don't have enough information"), while Style A's neutral tone felt less reliable in sensitive contexts. Style B showed stronger emotional awareness, with empathetic follow-ups such as "This sounds stressful, let's break it down step by step," whereas Style A often gave generic replies. A clear majority preferred Style B, describing it as "more human and less judgemental" and saying "I trusted its advice because it explained things clearly." Those who chose Style A valued its speed but admitted it lacked warmth. Overall, Style B consistently outperformed Style A, reinforcing that empathy and emotional intelligence are critical for healthcare AI.

**Task-based Usability Testing.** Task-based usability testing was conducted to evaluate LunaAI's performance across several usability metrics: task completion rate, time on task, error rate, clarity of prompts and feedback, and emotional comfort. Usability testing showed promising results, with all participants successfully completing core tasks such as navigating to the chatbot and asking health-related questions, adjusting chatbot tone, and accessing interactive features like voice input or uploading images.

The task completion rate was high, indicating that users could intuitively find and interact with the app's primary functions. The time on task was generally short, as most users could perform their intended actions without delays or confusion. Some errors were detected, such as the ambiguity of the "Get Started" button during onboarding, which was mistakenly perceived as swipe able. In terms of clarity of prompts and feedback, the chatbot provided easy to understand responses, respectful tone, and offered example prompts and follow-up suggestions, helping users proceed smoothly. Lastly, emotional comfort was consistently high, with users noting that the chatbot felt non-judgmental, calm, and supportive even during moments of emotional vulnerability. Overall, these outcomes reflect strong usability performance with room for minor refinements in visual cue design.

**Post-Task Surveys: The Functionality Ratings.**

Table 6. UI Functionality Ratings

| Feature | Rating (out of 5) | User Feedback Focus |
| --- | --- | --- |
| Login | 4.6 | Ease of completing the sign-up and login process |
| Chat History & Chatbot Interaction | 4.5 | Clarity and accessibility of previous chats and chatbot interactions |
| Notification | 4.5 | Helpfulness and ease of navigating reminder notifications |
| Profile | 4.8 | Ease of exploring and using features on the profile page |
| Settings | 4.7 | Ease of finding and adjusting necessary settings |

Table 6 shows that LunaAI's UI was rated highly across all features, with the Profile (4.8) and Settings (4.7) features leading in satisfaction, and Login, Notifications, and Chat History also scoring well (4.5–4.6). Users praised the app as smooth, accessible, and easy to navigate. Quantitative results further highlighted that most users found LunaAI polite, fair, and respectful, with many willing to use it again. Feedback emphasized its empathetic tone and clear, reassuring language, though suggestions included adding multilingual support and more personalized advice. Overall, the results show LunaAI effectively balances usability with emotional support, setting it apart from traditional healthcare chatbots.

Figure 6 shows the user evaluations of LunaAI's communication highlighted strengths as well as areas of concern. The chatbot was rated highly for fair and respectful responses (4.7) and for using polite, professional language (4.7), with users finding its tone suitable for sensitive healthcare contexts. It performed strongest in creating a supportive environment (4.9) and in avoiding judgmental language (4.6). However, concerns were raised around neutrality (4.5), where subtle biases were occasionally noted, and consistency (4.4), where some users observed minor contradictions. These findings suggest that while LunaAI is generally respectful and empathetic, improvements in neutrality and consistency are needed to ensure reliable and unbiased communication.

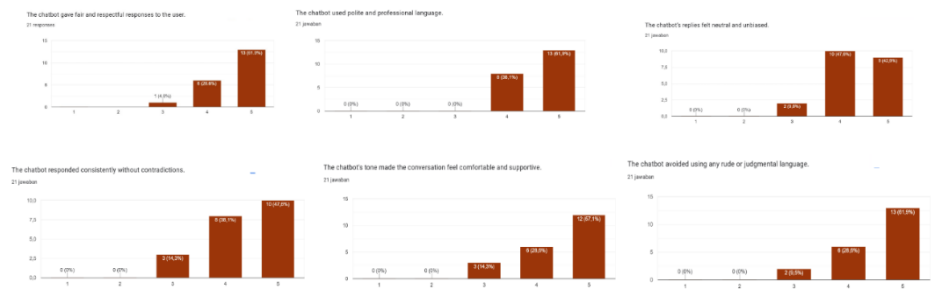

**Figure 6.** The user evaluations of LunaAI's communication

### 4.2    Heuristic Evaluation (Expert Review)

A mixed-methods technique that combined user testing and heuristic evaluation was used to verify the prototype. Following independent heuristic evaluations, four evaluators with backgrounds in healthcare chatbots—representing a range of ages, genders, and demographics- conducted group conversations. Ten modified heuristics, divided into fairness and politeness categories, served as the review's compass. The system was evaluated by the reviewers based on its ability to handle confusing or emotional inputs, respond consistently, handle tone and language sensitivity, and be transparent about its limits (such as its inability to diagnose). All things considered, the system received good grades for justice and civility, with particular strengths in clarity, empathy, and inclusivity. The reviewers assessed the system on: tone and language sensitivity, consistency in responses, handling of ambiguous or emotional inputs and transparency in limitations (e.g., not diagnosing).

**Fairness.** Figure 7 presents the heuristic evaluation analysis of fairness in LunaAI, assessed by five evaluators using severity ratings from 0 (not a problem) to 4 (usability catastrophe). The results indicate that LunaAI generally used inclusive and neutral language, respected cultural health practices, and offered information at different complexity levels, with most evaluators identifying no major issues. Minor concerns were noted in avoiding assumptions based on age or gender, providing sufficient user choices, and consistently acknowledging cultural sensitivities. These findings suggest that LunaAI demonstrates strong fairness in interaction design, with only small areas needing refinement to further support inclusivity and user autonomy.

More specific comments by the experts regarding fairness are as follows:

- Inclusive language: The chatbot steers clear of exclusionary terminology and instead speaks in a way that is friendly and impartial to all users.
- Avoiding age/gender assumptions: Answering solely to the information that the user has expressly supplied, without extrapolating demographic traits like age, gender, or social position.
- Providing options: Providing a variety of options or recommendations rather than strict directives or instructions to help consumers feel in charge of their choices.
- Respecting cultural health practices: Balancing evidence-based safety recommendations with traditional and cultural approaches to health without discounting them.
- Offering multiple complexity levels: Providing information that may be tailored to the needs of various users, from laypeople with basic explanations to those who want more in-depth knowledge.

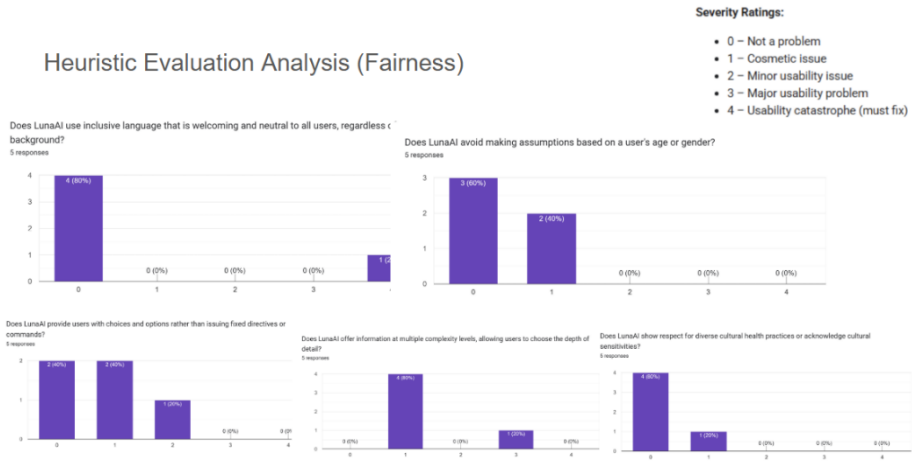

**Figure 7.** Heuristic evaluation analysis of fairness in LunaAI

**Politeness.** Figure 8 presents the heuristic evaluation analysis of politeness in LunaAI, assessed by five evaluators using severity ratings from 0 (not a problem) to 4 (usability catastrophe). The findings show that LunaAI consistently used respectful, gentle, and patient language and avoided judgmental or rushed tones, with all evaluators marking these as not a problem. The system was also noted to appropriately acknowledge user emotions and concerns, though a few evaluators suggested minor cosmetic refinements. Areas where slight usability issues were observed included avoiding unexplained jargon and consistently thanking users for their input. Overall, the analysis highlights politeness and respectful tone as clear strengths of LunaAI, with only minor areas identified for improvement.

More specific comments by the experts regarding politeness are as follows:

- Acknowledging emotions and concerns: Acknowledging and confirming users' emotions, especially when they show signs of anxiety, stress, or confusion.
- Respectful and patient language: Steer clear of a rushed or authoritarian delivery by using phrasing and tone that communicate encouragement, support, and composure.
- Avoiding unexplained jargon: Making technical phrases easier to understand so that individuals with different levels of health literacy can still access the information.
- Thanking users for sharing: Thanking users for sharing sensitive or personal information helps to build rapport and confidence.
- Avoiding judgment, rushing, or trivializing: Addressing every query with gravity and compassion; refraining from using words that downplay, disparage, or brush off customer concerns.

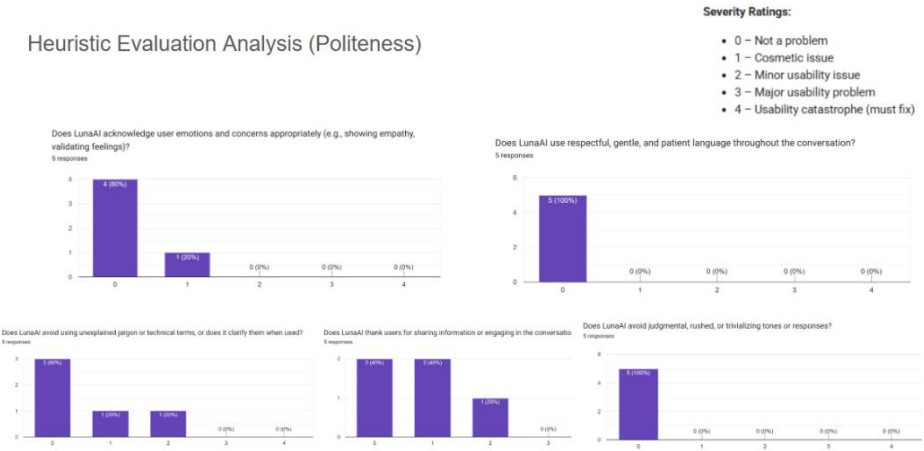

**Figure 8.** The heuristic evaluation analysis of politeness in LunaAI

### 4.3   Comparative Evaluation with Baseline Chatbots

To benchmark performance, LunaAI was compared to four widely available LLM chatbots: ChatGPT, Gemini, Claude, and Deep Seek. Each bot was given identical user prompts across three categories: health information guidance, emotional support, and symptom explanation.

**Table 7.** Comparison of LunaAI in Different AI Models -Based on user survey

| Bot | Politeness | Fairness | User Engagement |
|---|---|---|---|
| LunaAI | High | High | High |
| ChatGPT | Medium | High | Low |
| Deep Seek | Low | Medium | Low |
| Claude | Low | Low | Low |
| Gemini | Medium | Medium | Low |

Table 7 compares LunaAI with four widely available LLM-based chatbots—ChatGPT, Gemini, Claude, and DeepSeek—across politeness, fairness, and user engagement. The results show that LunaAI outperformed all other systems, achieving consistently high ratings in all three categories. Users highlighted LunaAI's ability to combine fairness and empathy with conversational warmth, leading to stronger engagement. In contrast, while ChatGPT demonstrated high fairness, its politeness was only rated as medium and engagement remained low, suggesting that factual accuracy alone did not create the same sense of trust and support. Gemini performed moderately in both fairness and politeness but still scored low in engagement, reflecting a more mechanical interaction style. DeepSeek and Claude scored lowest overall, with users describing their responses as clinical or lacking sensitivity, resulting in minimal engagement. These findings reinforce that social intelligence—expressed through fairness and politeness—plays a central role in sustaining meaningful user engagement in healthcare chatbots, and that LunaAI's design adaptations give it a clear advantage over general-purpose LLMs.

### 4.4   Input Handling

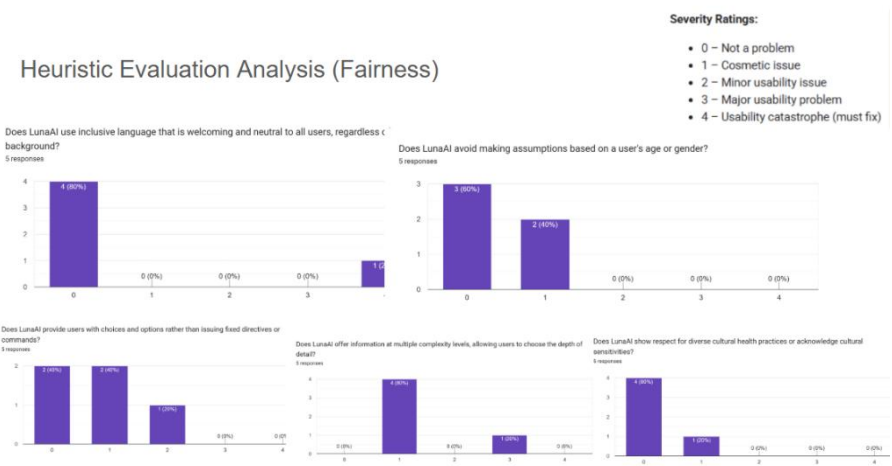

**Figure 9.** Visual Summary & User Feedback Patterns - Fairness

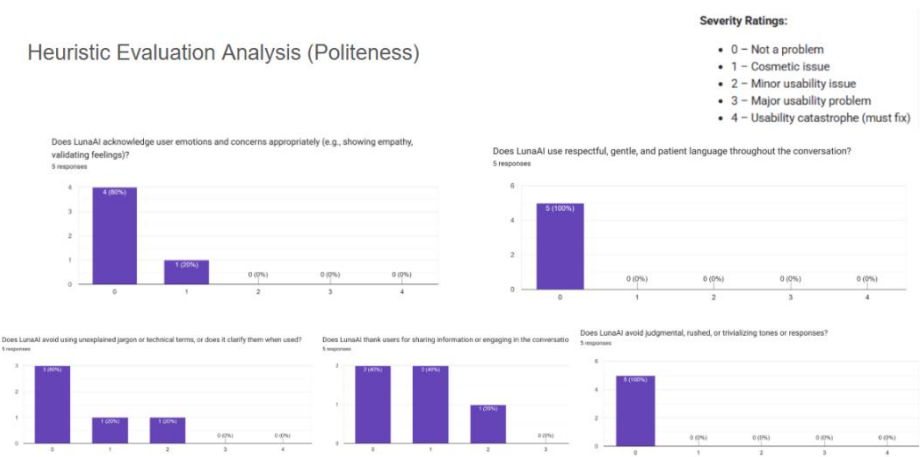

**Figure 10.** Visual Summary & User Feedback Patterns – Politeness

Figure 9 and 10 are the results of user feedback on Fairness and Politeness of LunaAI. The responses stood out for being more conversational, acknowledging user input, and offering follow-up questions. In contrast, other bots responded factually but lacked empathy or personalization. Example: When asked, "What should I eat to stay healthy?", LunaAI provided a supportive, flexible answer and invited the user to share more. ChatGPT gave a detailed breakdown but lacked warmth. DeepSeek was clinical, and Claude's response ignored context. The analysis also presented comparison charts rating system behavior across various metrics (politeness, clarity, helpfulness), with overall scores showing a medium-to-high performance across both expert and user ratings. Experts generally rated it medium on robustness but high on user control, while users rated it high on friendliness and response tone.

## 5    Discussion

The evaluation of LunaAI provides critical insights into the effectiveness and limitations of embedding fairness and politeness in conversational healthcare AI. Overall, users affirmed that interactions felt respectful, inclusive, and emotionally supportive, validating the hypothesis that these qualities are central to trust and acceptance in healthcare contexts rather than peripheral design features. Importantly, participants consistently preferred the polite and fairness-enhanced style (76%), describing it as "more human and less judgemental" and "like talking to a nurse," underscoring that emotional resonance can outweigh efficiency. The high politeness ratings (92% agreement) demonstrate the value of embedding Brown & Levinson's Politeness Theory [1] into conversational design. LunaAI's ability to maintain empathetic language while handling anxious or even hostile inputs shows that emotional intelligence can be consistently engineered into AI systems. This is notable given research showing that LLMs often default to negative politeness strategies, leading to pragmatic mismatches with human expectations [6]. By balancing positive face-saving (encouragement) with negative face-saving (user autonomy), LunaAI shows that explicit politeness frameworks can mitigate these limitations and enhance patient comfort, consistent with findings from Bowman Cooney et al. [5] that excessive formality can feel inauthentic in mental health contexts.

Fairness ratings (90% agreement) are equally significant in Malaysia's multicultural healthcare environment. LunaAI avoided exclusionary language, acknowledged traditional health practices without dismissing them, and offered multiple levels of explanation to suit diverse literacy needs. These findings align with prior concerns about algorithmic bias [7] and the challenges faced by non-dominant language speakers [22], suggesting that explicit fairness mechanisms can reduce disparities while still upholding evidence-based safety. The system's approach resonates with earlier arguments that developers bear an ethical obligation to design for inclusivity [8], while also addressing cultural sensitivity challenges in multilingual healthcare communication [21]. Expert heuristics confirmed strengths in inclusivity and tone sensitivity but flagged concerns around neutrality (4.5) and consistency (4.4), echoing literature on subtle biases in LLM-generated outputs [10], [11]. Nonetheless, challenges persist. LunaAI's responses to chronic or highly personalized health concerns sometimes lacked depth, reflecting the trade-off between safety constraints and personalization in healthcare AI noted in prior work on LLM limitations [13]. Similarly, while cultural sensitivity was demonstrated, fully dynamic multilingual and cross-cultural adaptation is still emerging, especially in Malaysia's diverse healthcare landscape. Suggestions from users — such as multilingual support, personalization for chronic conditions, and faster emergency access — highlight design implications that align with calls for culturally adaptable healthcare chatbots [23]. Taken together, these results show that fairness and politeness can be systematized through intentional design, prompt engineering, and heuristic-based iteration. LunaAI bridges the gap between technical competence and human-like social intelligence, offering evidence that healthcare AI can move beyond functional accuracy toward empathetic, trustworthy, and culturally sensitive companionship.

## 6      Limitations

LunaAI has several limitations that warrant consideration. Personalizing tone and responses proved challenging, as efforts to tailor communication risked reinforcing biases or undermining consistency. Achieving neutrality while adapting to individual preferences and diverse cultural norms of politeness and emotional expression also introduced risks of misinterpretation. The prototype was tested primarily within the Malaysian context, which limits generalizability without broader cross-cultural validation. Technical constraints restricted advanced capabilities such as real-time emotion detection, robust multilingual support, and seamless personalization, while the development timeline limited iterative refinement and feature depth. Practical healthcare integration was also out of scope, meaning the system's interoperability with medical providers, data governance standards, and regulatory compliance remain untested. Finally, both user testing and expert evaluations were conducted on a relatively small scale, which may have introduced sampling bias and limited the robustness of conclusions.

## 7      Conclusion & Future Work

This study introduced and evaluated LunaAI, a healthcare chatbot designed to embody fairness and politeness in its conversational interactions. The findings indicate that integrating ethical communication principles into conversational AI not only improves user satisfaction but also strengthens trust and usability, especially in emotionally sensitive environments like healthcare. By using a user-centered design approach, informed by politeness and fairness heuristics theory, LunaAI was able to significantly outperform generic LLMs in user engagement, emotional tone, and perceived fairness. The system's architecture, comprising prompt engineering, safety filters, and scenario-based design, offers a replicable model for future healthcare AI systems. Future work will focus on three primary areas:

1. Scalability and Real-world Deployment: Expanding user testing to larger and more diverse groups across public healthcare settings will help validate generalizability and inform practical integration strategies.
2. Multilingual and Cultural Adaptation: Deepening cultural sensitivity by incorporating dynamic, language-specific politeness norms and expanding natural language support across Malaysia's major languages.
3. Contextual Personalization: Enhancing long-term interaction memory and chronic condition support while maintaining fairness and safety, especially for users with low digital health literacy.

LunaAI represents a promising step toward more humane, respectful, and equitable healthcare AI. Its development reinforces the importance of ethical UX design in building systems that do not just answer questions but truly listen to users by keeping fairness and politeness as priorities.

## Appendix A:

**Links-Github**: https://github.com/novasingh/luna-bot